1 Chloroplast microsatellites: measures of genetic diversity and the effect of homoplasy

2 M. NAVASCUÉS and B.C. EMERSON

3 Centre for Ecology, Evolution and Conservation, School of Biological Sciences,

4 University of East Anglia, Norwich NR4 7TJ, UK

8 Correspondence: Brent Emerson, University of East Anglia, Norwich NR4 7TJ, UK.

9 Telephone: (44) 01603 592237. Fax: (44) 01603 592250. E-mail: b.emerson@uea.ac.uk

11 Running title: Chloroplast microsatellite homoplasy

**Abstract**

Chloroplast microsatellites have been widely used in population genetic studies of conifers in recent years. However, their haplotype configurations suggest that they could have high levels of homoplasy, thus limiting the power of these molecular markers. A coalescent-based computer simulation was used to explore the influence of homoplasy on measures of genetic diversity based on chloroplast microsatellites. The conditions of the simulation were defined to fit isolated populations originating from the colonization of one single haplotype into an area left available after a glacial retreat. Simulated data were compared with empirical data available from the literature for a species of *Pinus* that has expanded north after the last glacial maximum. In the evaluation of genetic diversity, homoplasy was found to have little influence on Nei's unbiased haplotype diversity ($H_e$) while Goldstein's genetic distance estimates ($D^2_{sh}$) were much more affected. The effect of the number of chloroplast microsatellite loci for evaluation of genetic diversity is also discussed.

**Introduction**

Microsatellites, or simple sequence repeats (SSRs), are sequences of repetitive DNA where a single motif consisting of one to six base pairs is repeated tandemly a number of times. Microsatellite sequences have been identified in the three eukaryote genomes: nucleus, chloroplast (Powell *et al.*, 1995) and mitochondria (Soranzo *et al.*, 1999) Nuclear microsatellite loci are usually highly polymorphic with alleles varying in the number of repeat units; they are codominant and inherited in a mendelian mode. These characteristics, plus being considered selectively neutral, have made them a popular marker for population genetic studies (Sunnucks, 2000). With respect to the organelle genomes, mitochondrial microsatellites have had little impact so far, but chloroplast microsatellites have been increasingly used in population genetics since their discovery. Conserved primers for the amplification of chloroplast microsatellites (cpSSRs) have been reported for conifers (Vendramin *et al.*, 1996), gramineae (Provan *et al.*, 2004) and dicotyledons (Weising & Gardner, 1999), but it is among conifers, for studies of population genetics, that chloroplast microsatellite markers have mainly been used (e.g. Cuenca *et al.*, 2003; Fady *et al.*, 2003; Gómez *et al.*, 2003).

Chloroplast microsatellites typically consist of mononucleotide motifs that are repeated eight to fifteen times. Levels of polymorphism in cpSSRs are quite variable across loci and across species, and some loci have been found to be monomorphic in all species studied. There are two important features that differentiate chloroplast from nuclear microsatellites. First, chloroplasts are uniparentally inherited. Some species have maternal inheritance of the chloroplast, others paternal. This means that cpSSRs provide

54   information for the lineages of only one of the sexes. Also, the chloroplast chromosome

55   is a non-recombinant molecule and, therefore, all cpSSR loci are linked. The genotyping

56   of cpSSRs will result in haplotypes that will be composed of the combination of alleles

57   found at each cpSSR locus.

58

59   Mutation rates for length variation in microsatellites have been found to be higher ($10^{-2}$ to

60   $10^{-6}$) than point mutations rates (Li *et al.*, 2002). In order to explain this difference two

61   kinds of mutational mechanism have been proposed: replication slippage (Tachida &

62   Iizuka, 1992) and recombination with out-of-phase aligning (Harding *et al.*, 1992). Both

63   processes result in changes in the number of repeat units which is compatible with the

64   observed size polymorphism of microsatellites. One consequence of these mutational

65   mechanisms is that the same genetic state (i.e. number of repeats) may evolve in two

66   different microsatellite lineages through independent mutational events, a phenomenon

67   known as homoplasy.

68

69   Homoplasy may cause problems in population genetic analysis as it can affect measures

70   of genetic diversity, gene flow, genetic distances (both between individuals and

71   populations), neighborhood size, assignment methods and phylogenetic analysis (see

72   Estoup *et al.*, 2002 for a review). Homoplasy within cpSSRs is considered as a potential

73   limitation for its use as a genetic marker (Provan *et al.*, 2001), however the problem has

74   only been addressed at the genus level (Doyle *et al.*, 1998; Hale *et al.*, 2004). Researchers

75   have generally considered homoplasy levels low enough to allow population genetic

76   analysis, and even when homoplasy has been evident (i.e. a haplotype network with up to

77    nine loops) it has been considered as "moderate" and its potential for confounding results

78    disregarded (Cuenca *et al.*, 2003).

79

80    In the present study, simulation analysis was used to investigate the evolution of cpSSRs

81    in conifers and how homoplasy may influence the informativeness of these markers.

82    Instead of a more traditional simulation approach where the whole population is

83    considered, only the genetic makeup of a sample of individuals was studied, following a

84    coalescence-based approach. The strategy consists of: first, generating a genealogy for a

85    sample of individuals from the last generation; and second, placing random mutations on

86    the genealogy to generate the genetic state of the sample (Hudson, 1990). This approach

87    has proved useful for the study of levels of homoplasy in nuclear microsatellites under

88    different mutational models (Estoup *et al.*, 2002).

89

**Modelling chloroplast microsatellite evolution**

*Simulation of Coalescent Events*

The probability of a coalescent event for any two lineages in a given generation depends on the population size, the population structure and the mating system. Thus, it became necessary to establish the biological scenario of the simulation, which would determine the shape of the genealogy.

After the last glacial maximum most conifer distributions shifted northward leaving their refugia (Jackson *et al.*, 2000). When populations were established, they had a period of expansion followed by a period of approximately constant population density until the present time (MacDonald & Cwynar, 1991). Our simulations reproduced populations under such conditions. A range of population ages (i.e. coalescence times) were investigated with population origins at 50, 100, 150, 200 or 250 generations before the present to simulate colonization events at different stages of the glacial retreat [assuming: 1 generation=100 years (as in Provan *et al.*, 1999) and the last glacial maximum=20 000YBP (Hewitt, 1996)]. Population growth was determined by the logistic equation:

$$N_{t+1} = N_t e^{r\left(1 - \frac{N_t}{K}\right)} \text{ (1)}$$

where $N_t$ is the population size in the generation $t$ ($N_0$=1), $r$ is the population growth rate and $K$ is the carrying capacity of the population. This model has been widely employed to describe population growth following colonization events (Shigesada & Kawasaki, 1997). The growth rate was set at $r$=0.7, producing a period of expansion with a duration of 22 generations (~2200 years), which is within the range observed for *Pinus* (MacDonald & Cwynar, 1991). Then, population size remained constant at carrying capacity for the

113      remaining 28, 78, 128, 178 or 228 generations. Because effective population sizes are

114      considered to be "large" in forest trees (Muona & Harju, 1989), carrying capacity was

115      arbitrarily set at $K$=10 000; a size big enough to avoid the effects of genetic drift

116      (Savolainen & Kuittinen, 2000).

117

118      Generally, chloroplasts are considered paternally inherited in conifers, although there is a

119      possibility that low levels of maternal leakage and heteroplasmy may be present (Cato &

120      Richardson, 1996). Thus, the coalescent process influencing the genealogy of chloroplast

121      haplotypes was considered to be the simplest case possible: a neutral, haploid, non-

122      recombinant genome with every individual having the same probability of being the

123      parent of any individual in the following generation. With the demographic history

124      determined, the genealogy was constructed using a generation-by-generation algorithm.

125      This type of algorithm allows the simulation of complex demographic and dispersal

126      models, contrasting with other coalescence algorithms which are faster for computational

127      time (Leblois *et al.*, 2003).

128

129      The algorithm worked backward in time, starting with the last generation and finishing in

130      the first one ($t$=0). In every generation the coalescent events were generated by an

131      algorithm that assigned to every individual, $x$, from the sample ($x \in [1,\ldots, n_t]$) in

132      generation $t$, its ancestor, $y$, in the previous generation $t$-1 (Fig. 1). Every individual $x$

133      from the sample in generation $t$ had a probability $P$=$n_{t-1}/N_{t-1}$ (where $n_{t-1}$ is the number of

134      ancestors already assigned and $N_{t-1}$ is the population size in generation $t$-1) to share its

135      ancestor $y$ with any of the individuals from the sample which ancestors had already been

136     assigned. A random number, $0<R<1$, drawn with a uniform probability distribution

137     function, is used to determine the occurrence of a coalescent event. When $R<n_{t-1}/N_{t-1}$, the

138     ancestor $y$ of the individual $x$ is within the $n_{t-1}$ previously assigned ancestors ($y \in [1,…, n_{t-1}]$).

139     In order to determine which ancestor, $y$ took the value of the integer part of $1+R \cdot N_{t-1}$

140     ($\mathrm{int}(1+R \cdot N_{t-1}) \in [1,…, n_{t-1}]$ when $R<n_{t-1}/N_{t-1}$). The generation and lineages involved in this

141     coalescent event were recorded to construct the genealogy of the sample. When $R>n_{t-1}/N_{t-1}$,

142     the ancestor $y$ was a new individual and $y$ took the value $n_{t-1}+1$; for the next individual

143     $x+1$ the value of $n_{t-1}$ increased by one unit.

144

145     Coalescent events were simulated using this algorithm on every generation until all

146     lineages converged to a single lineage. Because the population size in the first generation

147     ($t=0$) is only one individual ($N_0=1$) the probability for any number of individuals to share

148     their ancestors was $P=n_{t-1}/N_{t-1}=1$, i.e. all the lineages coalesced at least at the first

149     generation. This allowed controlling the coalescence time within the range of time for the

150     phenomenon simulated (i.e. colonization after glacial retreat). It is important to note that

151     due to the non-recombinant nature of the chloroplast genome all cpSSR loci were linked

152     and shared the same genealogical history.

153

154     *Simulation of Mutational Events*

155     Several theoretical mutational models have been proposed to describe microsatellite

156     evolution, each one of them with recognized weaknesses and strengths (Estoup &

157     Cornuet, 1999). All of them refer to, and have been tested against, nuclear microsatellites,

158     where mutational events are believed to occur by two mechanisms: replication slippage

159   and recombination (Li *et al.*, 2002). To our knowledge, no specific model has been

160   developed for cpSSRs (where no recombination occurs) so the stepwise mutation model

161   (SMM) was chosen since it is the simplest realistic model for microsatellites. Mutation

162   rate estimates for cpSSRs are scarce and vary from $10^{-3}$ (Marshall *et al.*, 2002) to $10^{-5}$

163   (Provan *et al.*, 1999) per locus per generation, so simulations were run under four

164   different mutation rates: $10^{-3}$, $5 \times 10^{-4}$, $10^{-4}$, $10^{-5}$ per locus, per generation.

165

166   The genetic state of nine cpSSR loci in samples of 25 individuals was simulated under 20

167   different combinations of population ages and mutation rates (Table 1). For each of these

168   simulations 20 replicates were run. The output of every replicate consisted of the

169   genotypic information of all the individuals from the sample and the genealogical tree

170   that describe their relationships, with information on the number of mutations and

171   number of generations for every branch. The raw data obtained was analyzed as

172   described in the following section.

173

**Genetic diversity analysis**

The effect of homoplasy was studied on a number of standard measures of genetic diversity for cpSSRs: total number of haplotypes, $N$ (direct count of different haplotypes); effective number of haplotypes, $N_e$ (reciprocal of the chance that two randomly chosen alleles are identical); unbiased haplotype diversity, $H_e$ (Nei, 1978), and average genetic distances among individuals, $D^2_{sh}$ (Goldstein $et\ al.$, 1995) applied to cpSSRs by Morgante $et\ al.$ (1998):

$$N_e = 1 \bigg/ \sum_{h=1}^{N} p_h^2 \quad (2)$$

$$H_e = \left[ \frac{n}{n-1} \left( 1 - \sum_{h=1}^{N} p_h^2 \right) \right] \quad (3)$$

$$D^2_{sh} = \frac{2}{[n(n-1)]} \cdot \frac{1}{L} \cdot \sum_{i=1}^{n} \sum_{j=i+1}^{n} d_{ij}^{\;2} \quad (4)$$

$$d_{ij} = \sum_{k=1}^{L} \left| a_{ik} - a_{jk} \right| \quad (5)$$

where $n$ is the number of individuals in the simulated sample, $p_h$ is the relative frequency of the $h^{th}$ haplotype, $N$ is the number of different haplotypes in the simulated sample, $L$ is the number of loci simulated, $a_{ik}$ is the size (measured in repeat units) of the allele for the $i^{th}$ individual and at the $k^{th}$ locus, and $a_{jk}$ is the size of the allele for the $j^{th}$ individual and at the $k^{th}$ locus.

Indices $N$, $N_e$ and $H_e$ were calculated both for the stepwise mutation model (SMM; where haplotypes were defined by their genetic state) and for the infinite allele model (IAM; where every mutation defined a new haplotype, even if the haplotype produced was

194 already present in the sample). The difference between the SMM and IAM values

195 represents information about genetic diversity that is lost due to homoplasy.

196

197 Estoup *et al.* (2002) defined an index of homoplasy, *P*, to quantify theoretically the

198 effects of mutational and population variables on homoplasy. This index of homoplasy is

199 the probability that two haplotypes sharing the same genetic state are not identical by

200 descent. We have calculated a similar index generated from Nei's genetic diversity for

201 the SMM and IAM:

202 $$P = 1 - \left( \frac{1 - H_{eIAM}}{1 - H_{eSMM}} \right) \quad (6)$$

203 The number of mutations occurring between every pair of lineages was scored and the

204 average genetic distance, based on number of mutations ($D^2_M$), was calculated following

205 Equation (4) where $d_{ij}$ is substituted for the number of mutations scored between the

206 individual *i* and *j*. Due to the possibility of recurrent mutation and back mutations under

207 the SMM, the genetic distance estimate is expected to be incongruent to some degree

208 with the actual number of mutations between lineages. The differences between the

209 absolute values $D^2_{sh}$ and $D^2_M$ were compared. In addition, the correlation of the matrices

210 of actual and estimated genetic distances was analyzed with a Mantel test (Mantel, 1967)

211 when the difference between $D^2_{sh}$ and $D^2_M$ was large.

212

**Results and discussion**

The results for all the simulations are presented in Figures 2 and 3. Each of the 20 replicates performed for each simulation is equivalent to an independent random sampling from the same population. Hence, for any of the genetic diversity indices, the mean value for the 20 replicates is interpreted as an estimate for the actual population value of that statistic and the standard deviation as the error associated to the sample size used. The effect of sample size was assessed by performing some simulations with larger sample sizes. Not surprisingly this resulted in a reduction of the variance for the different measures; for instance, gene diversity in simulation 16 ($H_{eIAM}$=0.552±0.178; $H_{eSMM}$=0.542±0.178 for 25 individuals, as shown in Figure 3G; $H_{eIAM}$=0.561±0.037; $H_{eSMM}$=0.553±0.036 for 1000 individuals) and average genetic distance in simulation 20 ($D^2_{sh}$=1.665±0.382; $D^2_M$=2.746±0.706 for 25 individuals, as shown in Figure 3H; $D^2_{sh}$=1.696±0.185; $D^2_M$=2.589±0.292 for 1000 individuals). The effect of the sample size could be eliminated by simulating the coalescent history for the whole population. However that would require excessive computational time, and it does not appear to be a problem warranting this.

As was expected, the simulations with parameters that produced higher genetic diversity also showed higher levels of homoplasy (see Table 1 and Fig. 4). For simulations with mutation rates higher than $10^{-4}$, homoplasy caused an underestimation, to different degrees, for the four diversity indices. However, Nei's haplotypic diversity values for SMM and IAM were very close in all simulations.

236 In order to understand the effect of homoplasy on Goldstein's genetic distance estimates,
237 a more complex approach is necessary. A difference between the values of $D^2_{sh}$ and $D^2_M$
238 will lead to an underestimation of the absolute time of coalescence for that sample.
239 However, the actual and estimated distances between individuals could be correlated, and
240 if that were the case, then the distance estimates could be used, or even corrected, to
241 study relative genetic distances. In order to test that correlation, simulation 20 was chosen
242 as it presents the biggest differences between estimated and actual distances. The Mantel
243 test performed for the correlation of the actual and estimated genetic distance resulted in
244 a significant correlation (*p-value* < 0.025) for all the replicates, with the correlation
245 coefficient, *r*, ranging from 0.52 to 0.82. Since the correlation with the actual genetic
246 distances is significant but not predictable, we conclude that, for populations with high
247 genetic diversity, the Goldstein's genetic distance estimates can be misleading, and any
248 attempt to apply a correction to these estimates would be prone to error.

249

250 The different effect of homoplasy on the various indices is explained by the nature of
251 these indices. The three indices based on the number of haplotypes ($N$, $N_e$ and $H_e$) seem
252 to perform better for assessing levels of genetic diversity than the one based on distance
253 estimates ($D^2_{sh}$). Estimated distances are influenced by every parallel mutation and back
254 mutation; however, these mutations may have no influence in the number of haplotypes
255 in the sample. For example, two lineages with unique and parallel mutations will have
256 two distinct haplotypes at the final generation, thus the number of haplotypes will not be
257 reduced. However, the genetic distance will be underestimated because of the parallel
258 mutations. Within the indices based on number of haplotypes, the indices that consider

259  their frequencies ($N_e$ and $H_e$) were less affected by homoplasy. This result is due to the

260  low frequency that most of the homoplasic haplotypes had within the simulated

261  populations.

262

263  In the present work, methods for phylogenetic reconstruction have not been assessed.

264  However, our results discourage the use of cpSSR to infer phylogenetic relationships. For

265  instance, in simulation 20, the high occurrence of homoplasic mutations (78.8% parallel

266  mutations and 6.4% back mutations) would make parsimony method useless. There

267  would be an effect on distance methods too, because distance estimates are affected by

268  homoplasy. However, further work would be necessary to estimate rates of error.

269

270  The number of cpSSR loci studied also influenced levels of homoplasy. A set of

271  simulations performed for four loci produced higher values for the homoplasy index, $P$,

272  and higher differences between expected and actual average distances than the equivalent

273  simulations performed for nine loci (Fig. 4). Thus, the linkage of cpSSRs can be seen as

274  beneficial for the analysis of genetic diversity. With a greater number of loci analyzed

275  one has more power to distinguish haplotypes with homoplastic alleles at a given cpSSR

276  loci through the polymorphism of linked loci.

277

278  Mutation rate and time of coalescence are the two factors influencing the levels of genetic

279  diversity our simulations. The different combinations of these parameters produced a

280  broad range of genetic diversity, from simulation 01 with null diversity to simulation 20

281  with the highest diversity (see Fig. 2 and 3). This set of simulations reproduces the

282  simplest scenario that could describe the recent history of a conifer population: an

283  isolated population originating from a single colonization event, followed by a population

284  expansion. The recent population history of *Pinus resinosa* would seem to be consistent

285  with such conditions. During the last glaciation *P. resinosa* was restricted to southern

286  refugial populations and has colonized northern areas after the glacial retreat (Fowler &

287  Morris, 1977). A study of the population genetics of *P. resinosa* with cpSSRs also

288  supports a metapopulation with restricted gene flow between populations (Echt *et al.*,

289  1998) This case provides us with an empirical study of a conifer with isolated populations

290  and with different colonization ages (either because of the metapopulation dynamics or

291  colonization after glacial retreat). Also, the number of cpSSR loci and sample sizes (nine

292  loci, 21-24 individuals) were similar to the simulations presented here, providing an

293  appropriate combination of conditions for comparison.

294

295  The patterns of genetic diversity found in the empirical study of *P. resinosa* with cpSSRs

296  (Echt *et al.*, 1998) can be best compared with simulations 11-15. In both cases the

297  populations are composed of one high frequency haplotype (the ancestral haplotype in the

298  simulations) plus several low frequency haplotypes. The diversity levels for the different

299  indices, $N$, $N_e$, $H_e$ and $D^2_{sh}$, are also comparable and consistent. Therefore, we can argue

300  that the genetic diversity and distances were unlikely to have been underestimated within

301  *P. resinosa* (see Fig. 2E, 2F, 3E and 3F).

302

303  Any further comparison of our simulations with other conifer species studied with

304  cpSSRs has to be done with caution. The simulations we have performed do not take into

305   account a number of additional factors that may influence homoplasy. Thus the current

306   set of simulations will represent the minimum amount of homoplasy that could be present

307   within a given population. Demographic scenarios including migration or more ancient

308   coalescent events (where colonization events were produced by more than one

309   haplotype), would result in increased levels of homoplasy. Higher mutation rates and size

310   constraints in the mutational model will also increase the levels of homoplasy (Estoup *et*

311   *al.*, 2002).

312

313   To conclude, further simulation studies would be beneficial for the understanding of the

314   homoplasy in the analysis of cpSSRs. In particular, we are now working in the

315   implementation of the generation-by-generation algorithm for the simulation of multi-

316   population scenarios with dispersal that will allow us to understand the effects of

317   homoplasy in the measurement of gene flow and genetic distances among populations.

318   Regarding future empirical studies with cpSSRs, it is strongly recommended that studies

319   use as many cpSSR loci as are available in order to reduce the negative consequences of

320   homoplasy on estimations of genetic diversity. In order to assess levels of genetic

321   diversity Nei's index seems to perform the best, being least affected by homoplasy. In

322   contrast conclusions made with Goldstein's genetic distances should be regarded with

323   caution, as these can underestimate absolute distances.

324

420 **Acknowledgments**

421 We wish to acknowledge the help of Guillermo de Navascués in the program

422 development. We are also grateful to Jo Ridley, Kamal Ibrahim and Godfrey Hewitt for

423 useful comments on early versions of the manuscript. We are grateful to the University

424 of East Anglia for the provision of a PhD studentship to MN.

425

426    **Fig. 1** Coalescence events were simulated with a generation by generation algorithm.

427    This algorithm assigned to every individual, $x$, of the generation $t$, its ancestor in the

428    immediately previous generation, $t$-1. The probability for this ancestor to be shared with

429    another individual was calculated from the population size in the previous generation, $N_{t-1}$,

430    $_{1}$, and the number of ancestors already assigned, $n_{t-1}$ (see text for details).

431

432 **Fig. 2** Number of haplotypes, *N*, and effective number of haplotypes, $N_e$, for the 20

433 simulations. Each graph represents the values for one of the indices, *N* or $N_e$, for the five

434 simulations with the mutation rate, $\mu$, shown on the left and the coalescence time (in

435 number of generations) shown on the abscissa axis. The mean and standard deviation

436 (from 20 replicates) is shown for each simulation. Both indices were calculated under the

437 stepwise mutation model (SMM) and the infinite allele model (IAM). The difference

438 between both values represents the extent to which information is lost due to homoplasy.

439

**Fig. 3** Unbiased haplotype diversity, $H_e$, and average genetic distances among individuals, $D^2$, for the 20 simulations. Each graph represents the values for one of the indices, $H_e$ or $D^2$, for the five simulations with the mutation rate, $\mu$, shown on the left and the coalescence time (in number of generations) shown on the abscissa axis. The mean and standard deviation (from 20 replicates) is represented for each simulation. $H_e$ was calculated for the stepwise mutation model (SMM) and the infinite allele model (IAM). $D^2$ was calculated for estimated distances ($D^2_{sh}$) based on the number of observed mutations and for the true distances based on the actual number of mutations ($D^2_M$). The difference between both values represents the extent to which information is lost due to homoplasy.

451  **Fig. 4** Levels of homoplasy plotted against genetic diversity for every replicate of the 20

452  simulations (see Table 1 for simulation conditions). Levels of homoplasy are represented:

453  (A) with the homoplasy index, $P$, and (B) with the difference between the actual average

454  genetic distance ($D^2_M$) and the estimated genetic distance ($D^2_{sh}$). Filled circles represent

455  simulations with nine loci and empty circles represent equivalent simulations with only

456  four loci.

457

458 **Table 1** Combinations of parameters for coalescence time and mutation rate used in
459 different simulations, and the index of homoplasy found for each.

| | Coalescence time [a] | Mutation rate, $\mu$ | Homoplasy index, $P$ [b] |
|---|---|---|---|
| Simulation 01 | 50 | $10^{-5}$ | 0.00 |
| Simulation 02 | 100 | $10^{-5}$ | 0.00 |
| Simulation 03 | 150 | $10^{-5}$ | 0.00 |
| Simulation 04 | 200 | $10^{-5}$ | 0.00 |
| Simulation 05 | 250 | $10^{-5}$ | 0.00 |
| Simulation 06 | 50 | $10^{-4}$ | 0.00 |
| Simulation 07 | 100 | $10^{-4}$ | 0.00 |
| Simulation 08 | 150 | $10^{-4}$ | 0.00 |
| Simulation 09 | 200 | $10^{-4}$ | 0.00 |
| Simulation 10 | 250 | $10^{-4}$ | 0.00 |
| Simulation 11 | 50 | $5 \times 10^{-4}$ | 0.01 |
| Simulation 12 | 100 | $5 \times 10^{-4}$ | 0.03 |
| Simulation 13 | 150 | $5 \times 10^{-4}$ | 0.03 |
| Simulation 14 | 200 | $5 \times 10^{-4}$ | 0.07 |
| Simulation 15 | 250 | $5 \times 10^{-4}$ | 0.17 |
| Simulation 16 | 50 | $10^{-3}$ | 0.02 |
| Simulation 17 | 100 | $10^{-3}$ | 0.06 |
| Simulation 18 | 150 | $10^{-3}$ | 0.11 |
| Simulation 19 | 200 | $10^{-3}$ | 0.33 |
| Simulation 20 | 250 | $10^{-3}$ | 0.43 |

460    $^a$ coalescence time in number of generations; $^b$ mean from 20 replicates

461

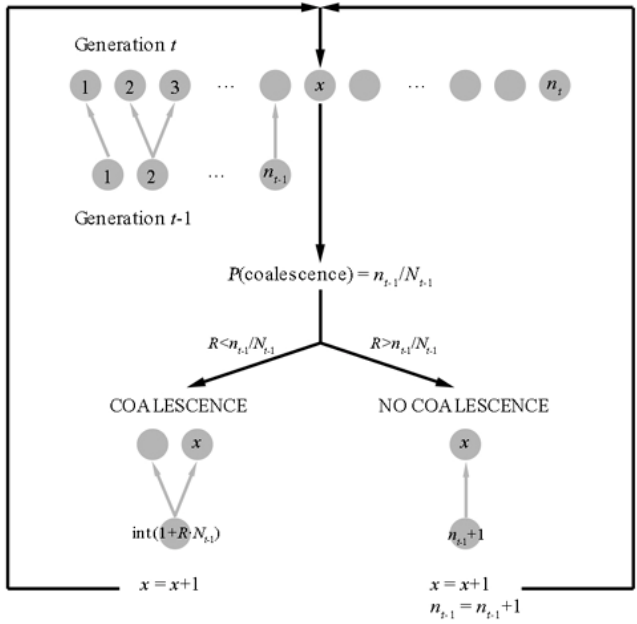

462

463

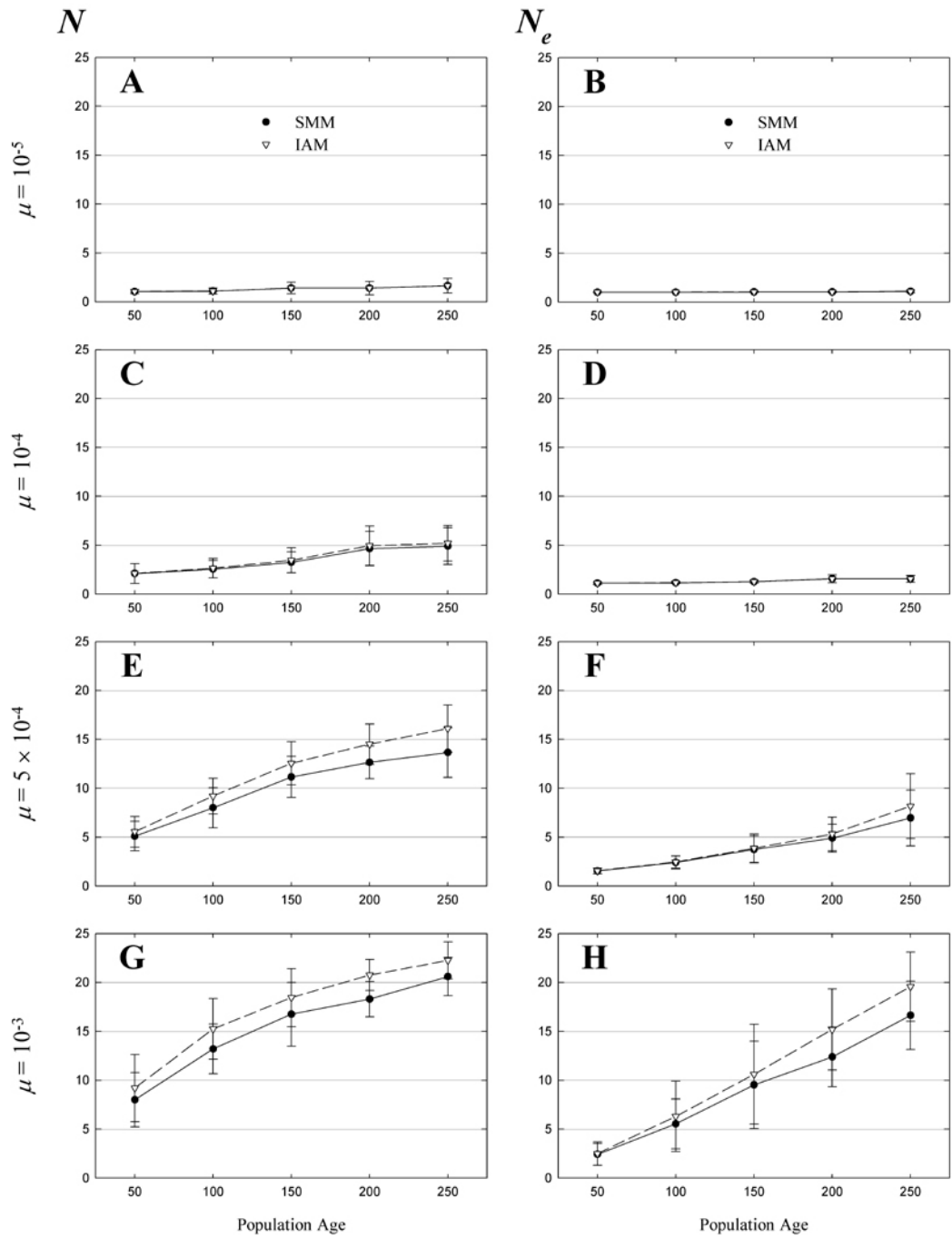

464

465

**$H_e$** $\qquad\qquad\qquad\qquad\qquad\qquad$ **$D^2$**

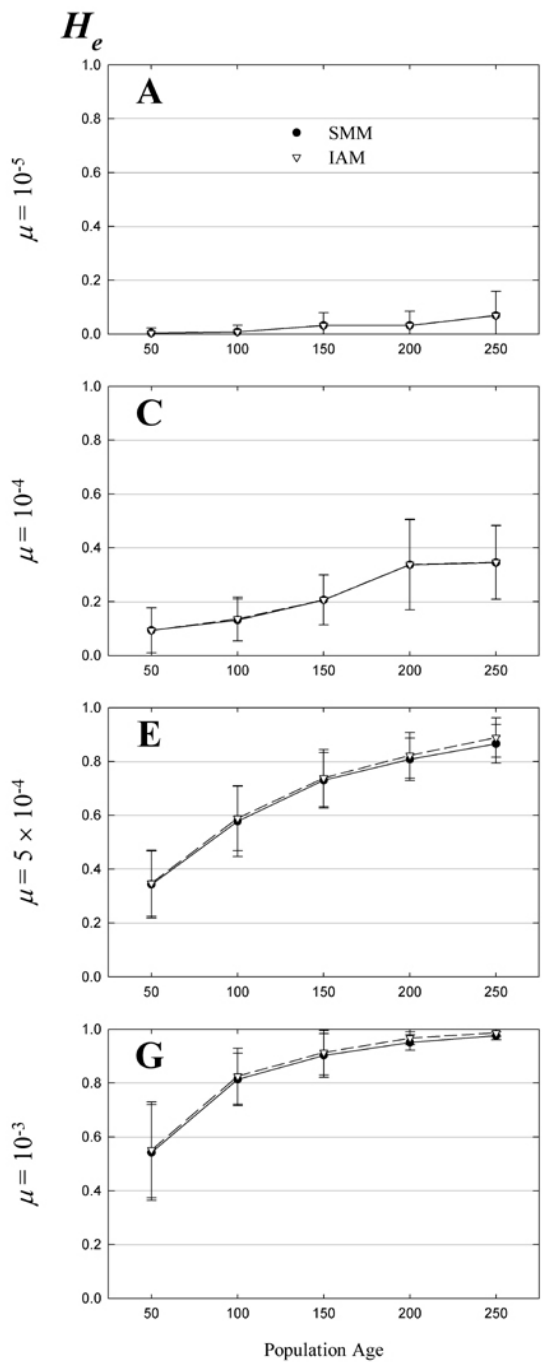
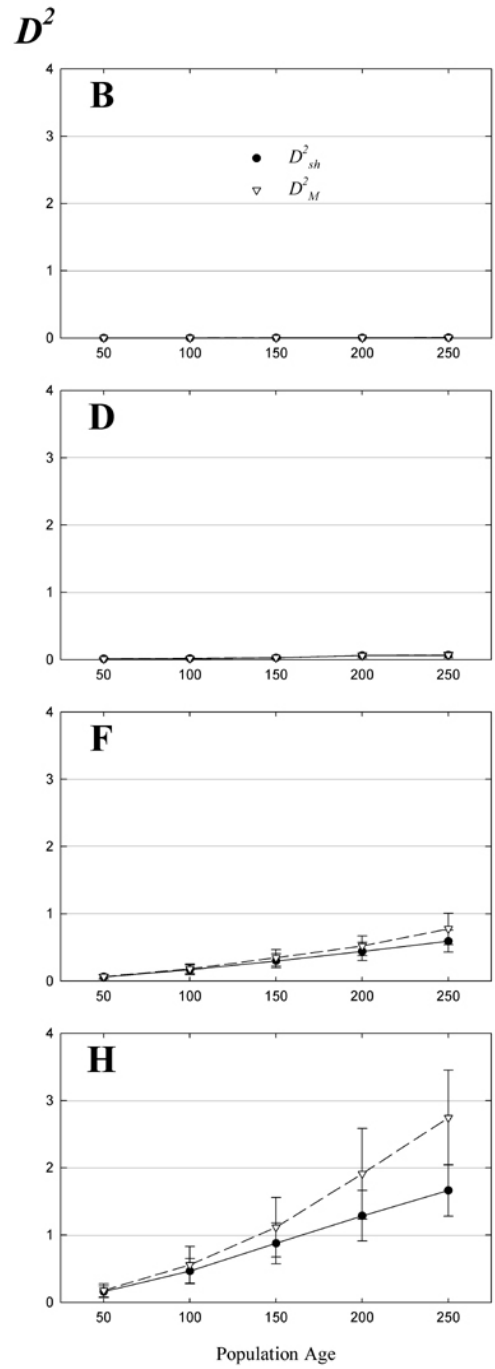

466

467

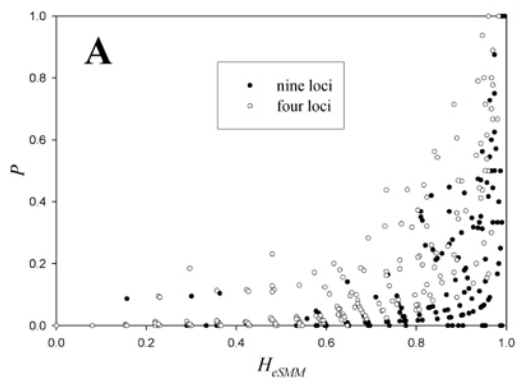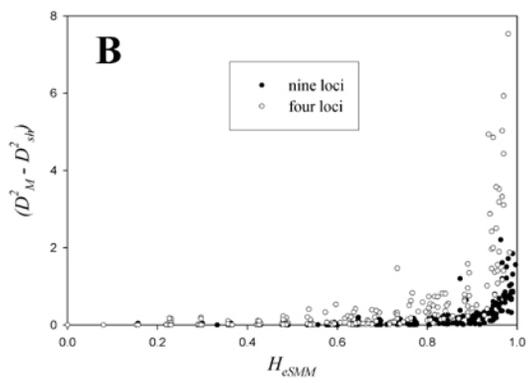

468